\documentclass[twocolumn,a4,fleqn]{svjour2}

\smartqed  
\usepackage{mathptmx}
\usepackage{graphicx}
\usepackage{hyperref}
\usepackage{amsmath}
\usepackage{bm}
\usepackage[sort&compress]{natbib}
\usepackage{graphicx}
\usepackage{doublespace}
\begin{document}
\title{A simple mechanism for controlling vortex breakdown in a closed flow.}

\author{Cecilia Cabeza \and Gustavo Saras\'ua \and Arturo C. Mart\'{\i}
\and Italo Bove}

\institute{ Cecilia Cabeza \email{cecilia@fisica.edu.uy} \and
Gustavo Saras\'ua \email{sarasua@fisica.edu.uy} \and Arturo C.
Mart\'{\i} \email{marti@fisica.edu.uy} \at Instituto de
F\'{\i}sica, Facultad de Ciencias,
  Universidad de la Rep\'ublica, Igu\'a 4225, 11400 Montevideo,
  Uruguay \and Italo Bove \email{italo@fing.edu.uy}
\at Instituto de F\'{\i}sica, Facultad de Ingenier\'\i a,
Universidad de la Rep\'ublica, J. H. y Reisig 565, Montevideo,
Uruguay }

\date{\today}
\date{Received: date / Accepted: date}

\maketitle

\begin{abstract}

This work is focused to study the development and control of the laminar
vortex breakdown of a flow enclosed in a cylinder. We show that vortex
breakdown can be controlled by the introduction of a small fixed rod
in the axis of the cylinder. Our method to control the onset of vortex
breakdown is simpler than those previously proposed, since it does not
require any auxiliary device system. The effect of the fixed rods may
be understood using a simple model based on the failure of the
quasi-cylindrical approximation.  We report experimental results of
the critical Reynolds number for the appearance of vortex breakdown
for different radius of the fixed rods and different aspect ratios
of the system. Good agreement is found between the theoretical and 
experimental results.


\end{abstract}

\section{Introduction}
\label{sec:int} The development of structural changes in
vortical flows and, particularly, vortex breakdown
\cite{Karman1921,Vogel1968,Leib78,Leib84,Escudier1984} has been
intensively investigated during the last years
\cite{Lopez1990a,Lopez1990b,Lopez95,Lopez96,Mark2003,Mullin2000,Mitchell2001,
Husain2003,Mununga2004,Fujimura2004,Piva2005,Lim2005,Zhang2006}.  The
characteristic and fundamental signature of vortex breakdown is the
appearance of a stagnation point followed by regions of reversed axial
flows with a bubble structure when the swirl is sufficiently large.
This structural change is also accompanied by a sudden change of the
size of the core and the appearance of disturbances downstream the
enlargement of the core. Vortex breakdown is very important in several
applications of Fluid Mechanics such as aerodynamics, combustion,
nuclear fusion reactors or bioreactors. The presence of vortex
breakdown may be beneficial or detrimental, depending on each
particular application \cite{Husain2003,Mununga2004}.

In order to explain the origin of vortex breakdown many proposals have
been given. Some of them are based on instability mechanisms, while in
other cases the proposed theories consider that hydrodynamical
instabilities do not play a significant role. In any case, it is
widely accepted that instabilities arise after the occurrence of
vortex breakdown.

Vortex breakdown (VB) has been observed not only in open flows, but
also in experiments performed in confined flows, for example, in
closed cylinders \cite{Escudier1984}. It is worth noting that the
characteristics of  VB in both cases are strongly similar,
suggesting the possibility that the basic mechanism of VB is the same
in both situations \cite{Mark2003}.

While considerable theoretical effort has been done in the study of VB
emergence in open flows, less results have been obtained for closed
flows.  In a way, this may be explained by the fact that in the later
case, the basic flow to be studied is much more complex than those
typically observed in open ducts. On one hand, experimental
measurements show that flows in open channels can be accurately
described with the relatively simple q-vortex model \cite{Leib84}.  On
the other hand, an analytical description for closed flows which are
considerable more complex than closed flows is not known
\cite{Escudier1984,Lopez1990a,Lopez1990b}.  However, from an
experimental point of view, experiments performed with flows confined
in cylinders are very attractive because they are simpler to control.
Due to the great number of practical implications of VB, the
development of mechanisms for controlling its emergence is of
considerable interest.

Recently, different methods for controlling VB in closed flows have
been proposed using different techniques such as co--rotation and
contour--rota\-tion of the end-walls \cite{Mununga2004}, the addition
of near-axis or at the end wall swirl \cite{Husain2003}, or
temperature gradients \cite{Herrada2003}.  Experiments show that such
methods may increase or decrease the critical Reynolds number to
develop VB.  In all the mentioned cases, it is necessary to use
auxiliary devices, such motors or heat sources.  The aim of this work
is to introduce a method simpler, from a practical point of view, than
those previously mentioned for controlling vortex breakdown in closed
flows.

This paper is organized as follows. In Sec.~\ref{model} we analyze a
simple model to predict the onset of vortex breakdown. In
Sec.~\ref{experiment} we present the experimental observations. These
results show that control of VB may be performed with the presence of
rods {\em at rest}, located along the axis of the cylinder, without
the necessity of transferring swirl. The comparison between the
experiments and the theoretical model is given in
Sec.~\ref{discussion}. Finally, in Sec.~\ref{summary} we present a
summary and the conclusions.

\section{The model}
\label{model} As already mentioned in the introduction, the
similarity between vortex breakdown in open and closed flows is
striking.  In particular, experiments and simulations show that, the
velocity fields around the point where VB develops, have the same
structure in open and closed flows.  The similarity of both situations
suggests that the formation of VB depends mainly on the local velocity
field, and the details of the flow far from the location of VB can be
neglected.  For example, in figure \ref{stream}, we can observe (a)
the experimental streamlines just before the appearance of VB from
Ref.~\cite{Escudier1984} and (b) a sketch of an open flow with
divergent section.  In subsection \ref{failure} we present a simple
model to predict the onset of VB in open flows.  We shall employ this
model to describe locally the enclosed flow and to study the influence
of local geometrical changes on VB. In particular, the effect on VB
development of a fixed rod located at the axis of the cylinder is
discussed in subsection \ref{effect-rod}. The results will be compared
with the experimental observations in subsequent sections.

\subsection{Failure of quasi--cylindrical approximation and VB}
\label{failure}
\begin{figure}
\centerline{
\includegraphics[width=8cm]{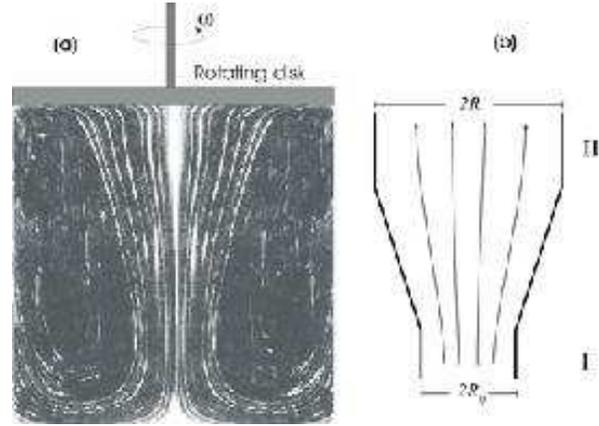}
}
\caption{(a) Streamlines of the enclosed flow prior to the development
of VB (taken from Ref. [Escudier 1981 ]).  The streamlines are
divergent near the point where VB develops, suggesting that locally
the flow may be described as the flow inside a duct that experiment an
expansion as it is shown in (b).}
\label{stream}
\end{figure}

In the past, it has been shown that simple theoretical vortex models
of open flows may predict behaviors very similar to those occurring in
vortex breakdown.  Various authors associated the emergence of VB with
the failure of the quasi-cylindrical approximation in inviscid
theories \cite{Leib84}. This approach gives a prediction about the
position of VB that is in good agreement with experiments
\cite{Leib78}. A classical approach based on an axisymmetric inviscid
analysis was given by Batchelor \cite{Batch}. In this model, that can
be applied outside the boundary layer, it is assumed that the
fluid in steady motion passes through a transition from a duct of
radius $R_{0}$ to another duct of different radius $R$ as it is shown in
Fig.~\ref{stream}(b).

Based on Batchelor's model, we assume that upstream, region I in
Fig.~\ref{stream}(b), the fluid moves inside a duct of radius $R_{0}$\,
following a Rankine vortex with uniform axial velocity.  This vortex
has a core of radius $c_0$ which rotates like a solid body with
constant angular velocity $\sigma$ and an outer irrotational region.
The whole vortex has an axial motion with velocity $U_0$.  In a
cylindrical system of coordinates ($r,{\theta},z$) with the z-axis in
the direction of the duct the velocity field is written as:
\begin{eqnarray}
\label{f1}
\ v_{r}&=&0, \\ \nonumber
v_{\theta }& =&\begin{cases}
             \sigma r  & 0 < r < c_0, \\
              \Gamma /r & c_{0}<r<R_{0},
          \end{cases}  \\
 v_{z}&=&U_{0}. \nonumber
\end{eqnarray}
where $\Gamma$ is the circulation in the outer region and in order to
assure the continuity of the velocity at $r=c_0$ it must satisfy $\Gamma
=\sigma c_{0}^{2} $.

Now, in order to solve the fluid motion downstream, region II in
Fig.~\ref{stream}, we introduce the axisymmetric streamfunction
$\Psi$ related to the velocity field by means of
\begin{eqnarray}
 v_{z}&=&\frac1r \frac{\partial \Psi}{\partial r}, \\
v_{r}&=& - \frac 1r \frac{\partial \Psi}{\partial z}.
\end{eqnarray}
From Euler's equation  the following equation for
the streamfunction is obtained \cite{Batch}:
\begin{equation}
r\frac{\partial }{\partial r}(\frac{1}{r}\frac{\partial \Psi
}{\partial r})+ \frac{\partial ^{2}\Psi }{\partial
z^{2}}=r^{2}\frac{dH}{d\Psi }-K\frac{dK}{ d\Psi }  \label{eqphi}
\end{equation}
where $H=\frac{1}{2}v^{2}+\frac{p}{\rho }$ , $K=rv_{\theta }$,
$\rho$ is the density, $p$ is the pressure and $v^2$ is the magnitude of
the velocity. For the flow given by Eqs.~(\ref{f1}), we have that
\begin{equation}
H=\frac{2\sigma ^{2}}{U_{0}}\Psi
+\frac{1}{2}U_{0}^{2},\;\;K=\frac{2\sigma }{ U_{0}}\Psi ,\; for \;
0<r<c_0  \label{relphi}
\end{equation}

Thus, Eq. (\ref{eqphi}) takes the form
\begin{equation}
 \Psi_{rr} -
\frac{1}{r}\Psi_{r}+ \Psi_{zz}=-\frac{4\sigma ^{2}}{U_{0}^{2}}%
\Psi +\frac{2\sigma ^{2}}{U_{0}} r^2  \label{eqphi2}
\end{equation}
where the subscripts, $r$ and $z$, denote derivatives.
The relations (\ref{relphi})  hold also downstream for the steady
flow, so that Eq. (\ref{eqphi2}) determines the cylindrical flow
downstream. The general solution for the downstream rotational
region of the flow for any kind of cylindrical ducts is
\begin{equation}
\Psi (r)=\frac{1}{2}U_{0}r^{2}+AF_{1}(\gamma r)+BY_{1}(\gamma r)
\label{solphi}
\end{equation}
where \ $F_{1}$\ and $Y_{1}$ are the Bessel functions of the first
and second kind respectively and  $\gamma = 2  \sigma / U_0$
\cite{Batch}. The downstream flow in the  irrotational region is
given by

\begin{eqnarray}
v_{r} &=& 0, \\
v_{\theta }&=& \Gamma /r, \\
v_{z}&=&U,
\end{eqnarray}
where $U$ is a constant that can be determined using conservation
laws.

Now, using the fact that downstream the fluid is constrained to move
inside a cylindrical duct of radius $R$, in order to avoid the
divergence of $\Psi $ at $r=0$ it follows that $B=0$. The no mass
flow condition at the solid boundary is automatically satisfied
because the flow has no radial component.  Therefore, the downstream
cylindrical flow in the rotational region ($r < c$, where $c$ is the
radius of the rotational core) is given by
\begin{eqnarray}
 v_{r}&=&0, \nonumber \\
v_{\theta } &=&\sigma r+\frac{A S}{R}  J_{1}(\frac{S}{R} r), \nonumber\\
v_{z}&=& U_{0}+\frac{A S}{R} J_{0}(\frac{S}{R} r), \label{solII}
\end{eqnarray}
whereas in the irrotational region ( $c < r< R$),  the flow is
written as
\begin{eqnarray}
v_{r}&=&0, \nonumber \\
v_{\theta } &=&\Gamma /r, \nonumber \\
v_{z}&=& \frac{R_0^2-c_0^2}{R^2-c^2} U_0,
\end{eqnarray}
where, using mass conservation, \begin{equation}
A=\frac{(c_{0}^{2}-c^{2})U_{0}}{2c J_{1}(\frac{S}{R}
c),}
\end{equation}
and $S$ is the swirl parameter, defined as $S= R \gamma = \frac{2
\sigma R}{U_{0}}$.  Using the continuity of the pressure, we obtain
the following implicit equation that gives the value of the core
radius $c$,
\begin{equation}
\frac{(R_{0}^{2}-c_{0}^{2})}{(R^{2}-c^{2})}-\frac{\frac{S}{R}
(c_{0}^{2}-c^{2})J_{0}(\frac{S}{R}
c)}{2c J_{1}(\frac{S}{R} c)}=1. \label{detr}
\end{equation}

\begin{figure}
\centerline{
\includegraphics[width=4in]{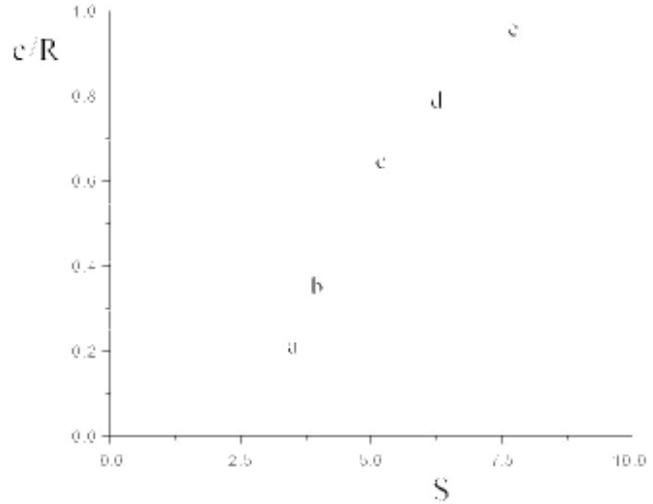}
}
\caption{Theoretical results for the dimensionless core size, $c / R$,
as a function of the swirl, $S$, and different radius of the fixed
rod, (a) $d=0$, (b) $d=0.2$, (c) $d=0.4$, (d) $d=0.6$ and (e) ,
$d=0.8$. Other parameter values: $R_0=1$, $c_0=0.25$ and $R=1.2$ }
\label{ratio1}
\end{figure}

We solved this equation for different values of the parameters.  When
the value of $S $ is increased, it may happen that two branches of
solutions for $c$ collide and disappear.  In Fig.~\ref{ratio1} the
curve labelled (a) represents the ratio $c/R$ as a function of the
swirl $S$. We observe in this figure that for each curve
(corresponding to the different values of $d$, parameter to be introduced
in the following subsection) there exists a critical value, above
which the solution does not exist.

In Batchelor approach \cite{Batch}, the disappearance of these
solutions is interpreted as the signal of VB emergence. We discuss now
a way to support this interpretation. As already mentioned, the
typical signature of  VB is the formation of a stagnation point,
followed by regions with reversion of the axial velocity.  In the
neighborhood of the stagnation point, a quasi-cylindrical description
of the flow is doomed to failure, because in this place the flow is
strongly dependent on the axial coordinate $z$ \cite{Leib78}.  This is
revealed by the disappearance of cylindrical solutions. In the absence
of VB a quasi-cylindrical description is suitable and, probably, there
exists a cylindrical solution to describe the flow.  Then, it is
reasonable to assume that the critical value $ S_{VB}$ for the
emergence of VB \ and the critical value $S_{c}$\ for the
disappearance of cylindrical solutions are very close to one another,
i.e. $ S_{VB}\approx S_{c}$, and the identification of both may be a
practical criteria for VB. From now on, we use this criterion to
\emph{estimate} the critical conditions for the emergence of VB. We
shall employ the above model of open flows to study qualitatively the
flow in the closed cylinder.  We shall assume that, if the open and
closed flows are similar around the localization of VB, then the
phenomena that takes place in both cases are similar. The flow inside
the cylinder is analogous to the open flow near a transition between
two cylindrical ducts (regions I and II of Fig.~\ref{stream}). We
assume that in region I the flow is given by Eq.~(\ref{f1}) and the
cylindrical flow in region II is given by the Eqs.~(\ref{solII}). Thus
the radius of the core $c$ is to be determined with the
Eq. (\ref{detr}). As already mentioned, for some range of values of $
R_{0}$, $c_{0}$ and $U_{0}$, cylindrical solutions do not exist in
region II if $S$ is above a critical value $S_{c}$ (see
Fig. \ref{ratio1}).  According to Batchelor's criterion, this means
that VB takes place inside region II, when the swirl parameter $S$ is
larger than a critical value $ S_{c}$. This behavior is in qualitative
agreement with VB inside the closed cylinder.

\subsection{Effect of rods presence at the axis}
\label{effect-rod}
We now estimate how the changes in the geometry of the region II
affect the development of VB. More specifically, we consider the
effect produced by the presence of cylinders along the duct
axis. Taking into account the local similarity between open and closed
flows, we extend the model introduced in the previous sub-section to
this case. Using the formalism described above, we obtain the
following equation that determines the radius of the rotational core
$c$ in presence of the inner cylinder of radius $ d$
\begin{equation}
\frac{(R_{0}^{2}-c_{0}^{2})}{(R^{2}-c^{2})}-\frac{A_{II}}{U_{0}}%
\frac{S}{R} J_{0}(\frac{S}{R}
c)-\frac{B_{II}}{U_{0}}\frac{S}{R} Y_{0}(\frac{S}{R} c)=1,
\end{equation}
with
\begin{equation}
A_{II}=\frac{U_{0}}{2cd}[\frac{d(c_{0}^{2}-c^{2})Y_{1}(%
\frac{S}{R} d)+c d^{2}Y_{1}(\frac{S}{R}
c)}{J_{1}(\frac{S}{R} c)Y_{1}(\frac{S}{R}
d)-J_{1}(\frac{S}{R} d)Y_{1}(\frac{S}{R} c)}],
\end{equation}

\begin{equation}
B_{II}=\frac{U_{0}}{2cd}[\frac{d(c_{0}^{2}-c^{2})J_{1}(%
\frac{S}{R} d)+c d^{2}J_{1}(\frac{S}{R}
c)}{Y_{1}(\frac{S}{R} c)J_{1}(\frac{S}{R}
d)-Y_{1}(\frac{S}{R} d)J_{1}(\frac{S}{R} c)}].
\end{equation}

\begin{figure}
\center
\includegraphics[width=4in]{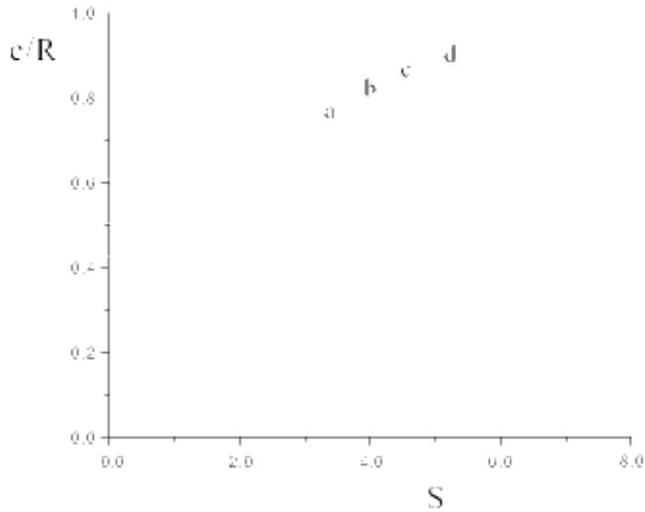}
\caption{As figure \ref{ratio1} but with $R_0=1$, $c_0=0.4$ and
$R=1.2$, for (a) $d=0$, (b) $d=0.2$, (c) $d=0.4$, (d)
$d=0.6$. }
\label{ratio2}
\end{figure}

\begin{figure}
\center
\includegraphics[width=4in]{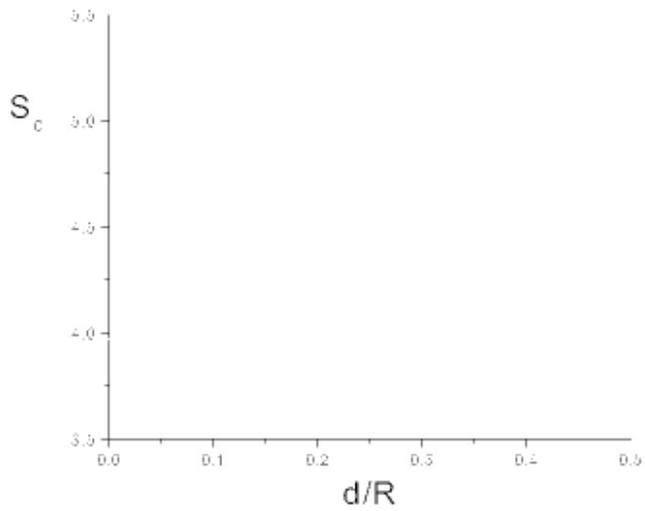}
\caption{Critical value of the
swirl parameter, $S_c$, for the disappearance of cylindrical solutions as a
function of $d/R$, for the case $R_0=1$, $c_0=0.4$ and $R=1.2$}
\label{Scritico}
\end{figure}

In order to appreciate the influence of the inner cylinder on the
disappearance of cylindrical solutions, figures \ref{ratio1} and
\ref{ratio2} show the ratio $c/R$ as a function of the swirl $S$. The
effect of a very slender inner cylinder may be to increase
(Fig.~\ref{ratio1}) or decrease (Fig.~\ref{ratio2}) slightly the
critical value $S_{c}$.  However, for all the situations considered,
the critical value $S _{c}$ increases as long as $d$ is above a
threshold.  In this case, the emergence of VB was transferred to
larger values of the swirl parameter, and VB was suppressed in the
range between the old and new critical values of $S$ (see
Fig.~\ref{Scritico}).  These results show that the model predicts the
suppressing effect of the slender cylinders. In the next section, we
present the experimental results of that were performed to study the
effect of the rods in confined flows.

\section{Experimental Setup and Results}
\label{experiment} 
\begin{figure}
\center
\includegraphics[height=5cm]{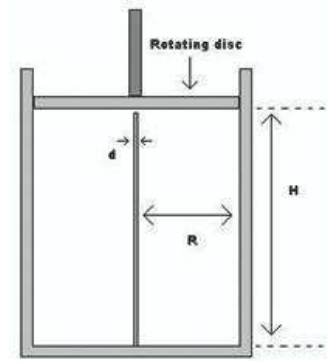}
\caption{Experimental setup. Closed cylinder, radius $R$ and height
$H$, with a rotating top wall. Along the axis of the cylinder a fixed
rod of radius $d$ is located.}
\label{setup}
\end{figure}

The experimental setup consists of an acrylic
cylindrical container of inner radius $R=40$ mm and a rotating top
disk at a variable height $H$ rotating with angular velocity
$\Omega$ (figure \ref{setup}). The fluid used was water
dissolutions of glycerin at 60\% in mass, with $\nu = 1 \times
10^{-5}$ m$^2$/s.  Temperature was kept constant at $20^oC$. The
Reynolds number corresponding to the rotating top wall, $Re=\Omega
R^2/ \nu$, varies between 600 and 2600, with a 1\% error. Four
different aspect ratios $H/R$ were used: 1, 1.5, 2 and 2.5.  The
visualization system consists of a vertical sheet of light with 2
mm of thickness, generated by two slide projectors. Small
quantities of fluorescein were injected into the flow through a
small hole in the bottom disk.  Pictures were taken using a 5
megapixel Canon digital camera.  To investigate the effect of the
inner cylinders on  VB, we have used three axial fixed rod of
radius $d=1$ mm, $d=2.5$ mm and $ d =5$ mm.

\begin{figure}
\center \includegraphics[height=6cm]{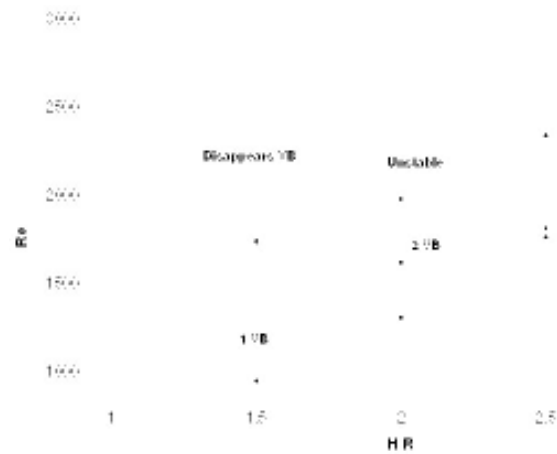}
\caption{Critical Reynolds number corresponding to the appearance of a
single VB (circles), double VB (squares), oscillatory bubbles
(diamond) and disappearance of VB (stars) for the case without fixed
rod.}
\label{stability}
\end{figure}

\begin{figure}
\center \includegraphics[height=5cm]{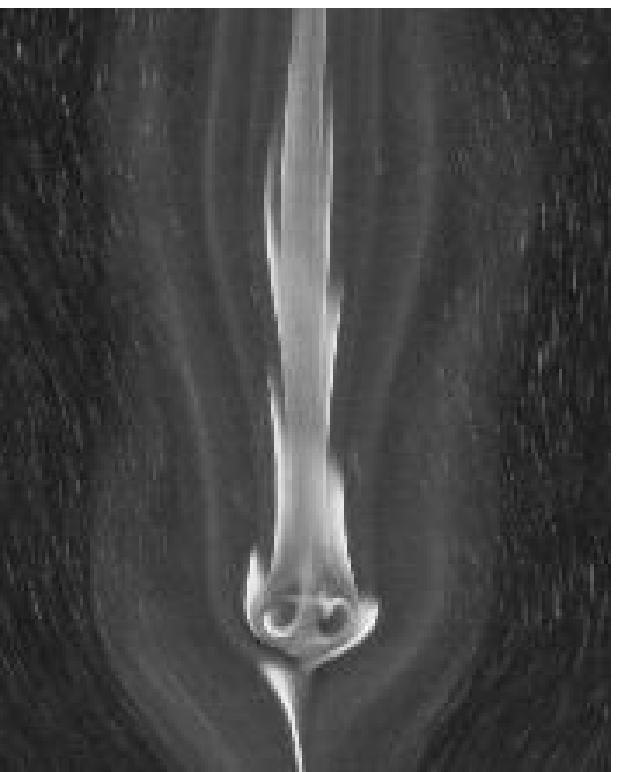}
\includegraphics[height=5cm]{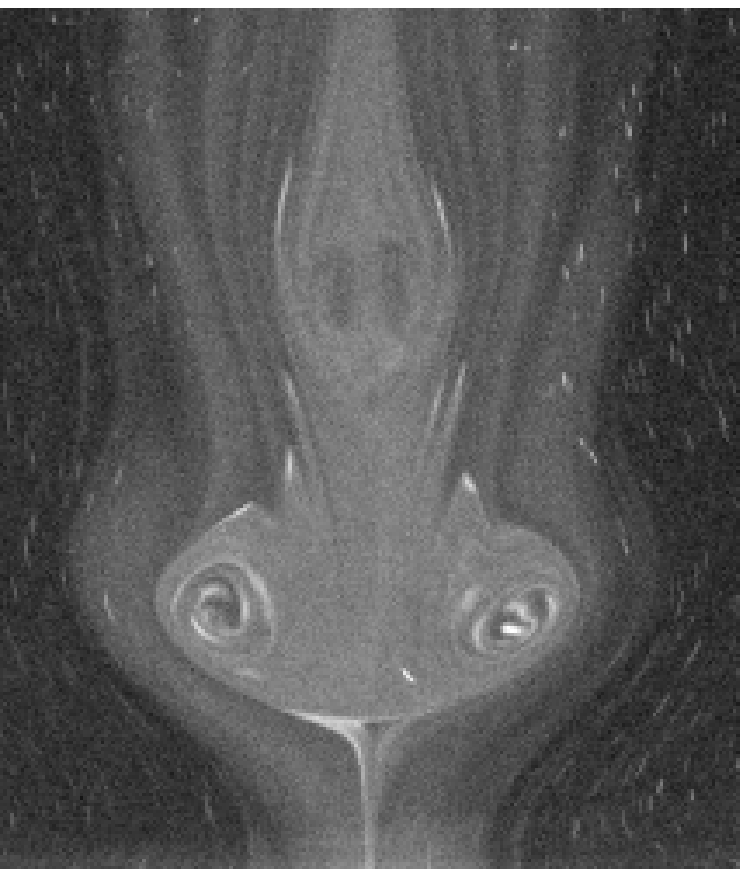}
\caption{Flow visualization showing one and two vortex breakdown,
without axial rod. (a) $\mathrm{Re}=1300$,
H/R = 2. (b) {\rm Re} = 1756, H/R=2.5.}
\label{sinbarra}
\end{figure}

For convenient comparison with previous works \cite{Escudier1984} , we
first studied the dynamical behavior of usual vortex breakdown without
axis rod. In Fig. \ref{stability} the experimental results for
different aspect ratios can be observed.  For $H/R = 1$, vortex
breakdown does not appear. For $H/R =1.5$, vortex breakdown take place
for ${\rm Re} = 940$. With further increment of Reynolds number, the
recirculation bubble becomes oscillatory (${\rm Re} = 1732$), and for
${\rm Re} = 1852$, it disappears.  These results agree with classical
work of Escudier \cite{Escudier1984}.  For $H/R = 2$ and 2.5, first
one VB, and subsequently two VB, are generated when the Reynolds
number is increased.  It is possible to see in figure \ref{sinbarra}
the interior detailed structure of the recirculation bubble.

\begin{figure}
\center \includegraphics[height=6cm]{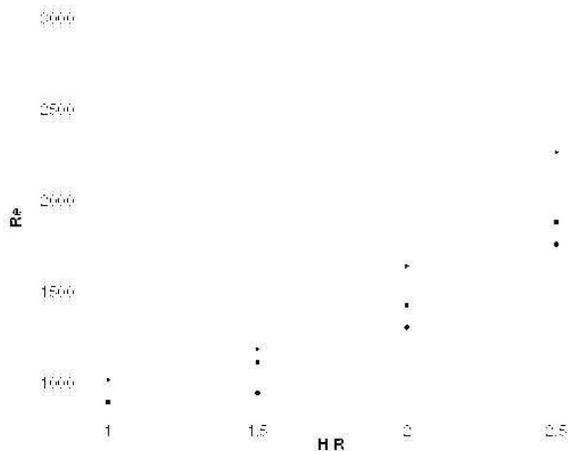}
\caption{ Critical Reynolds numbers corresponding to the appearance of the
first bubble as a function of the aspect ratio $H/R$. Without axial
rod (circles) and with axial rod; $d = 2.5$ mm (squares) and $d =
5$ mm (triangles).}
\label{diag}
\end{figure}

\begin{figure}
\center \includegraphics[height=7cm]{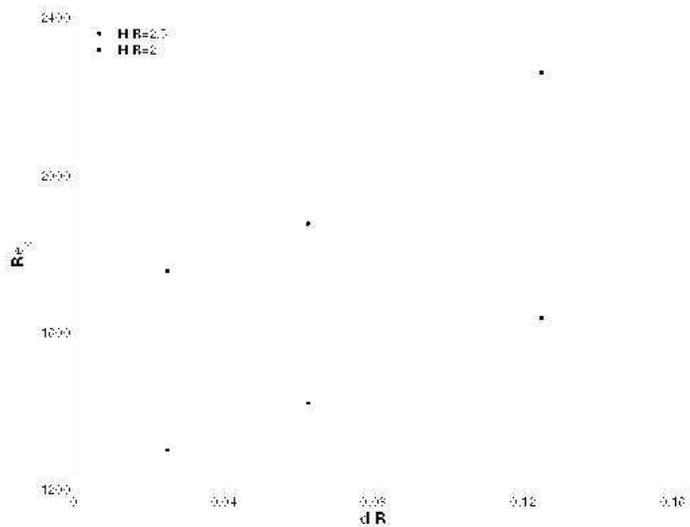}
\caption{Critical Reynolds number as a function $d/R$ for the
appearance of single VB and different aspect ratios, $H/R=2$ (squares)
and  $H/R=2.5$ (circles).}
\label{rec}
\end{figure}

\begin{figure}
\center
\includegraphics[height=5cm]{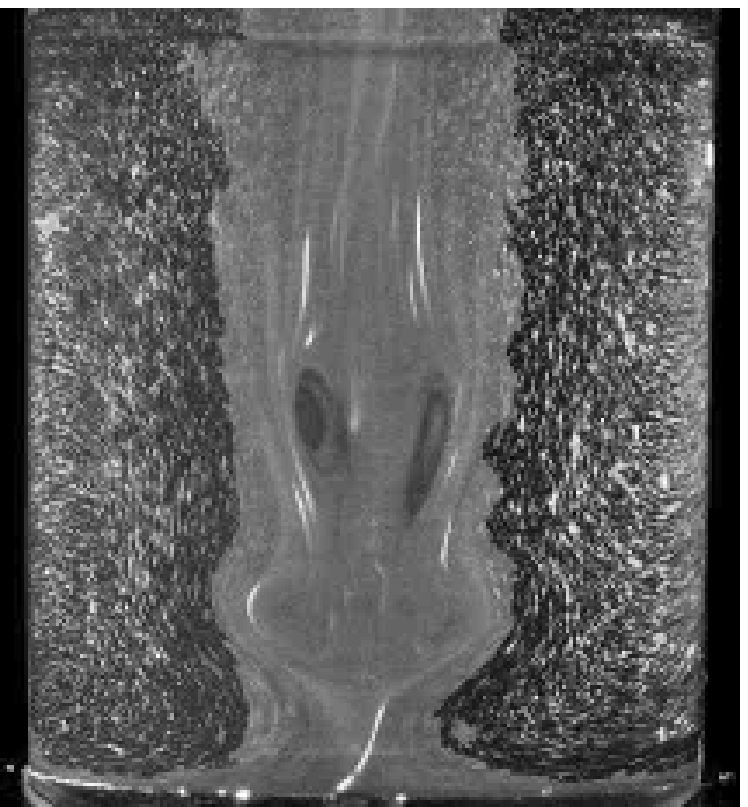}
\includegraphics[height=5cm]{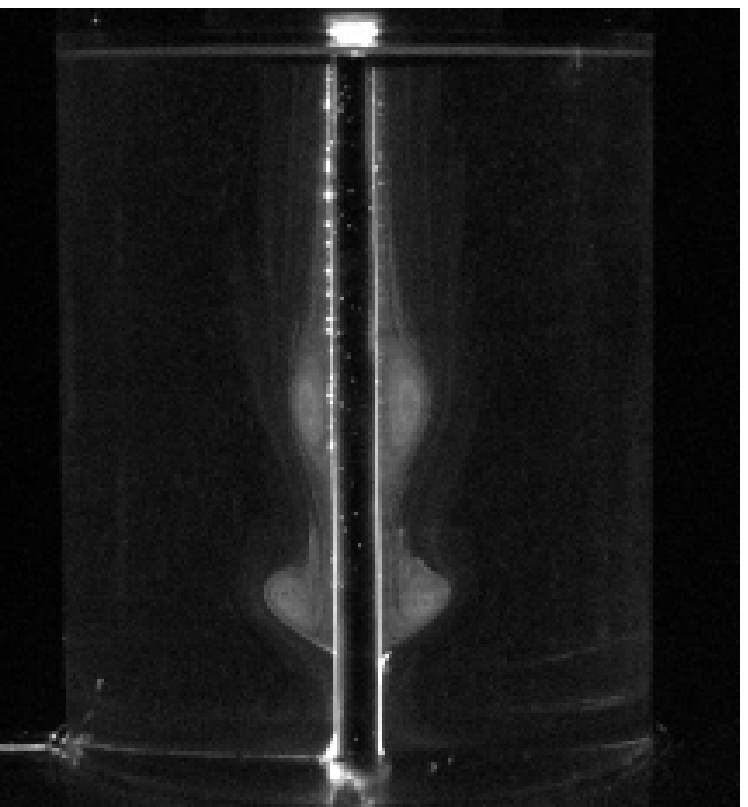}
\includegraphics[height=5cm]{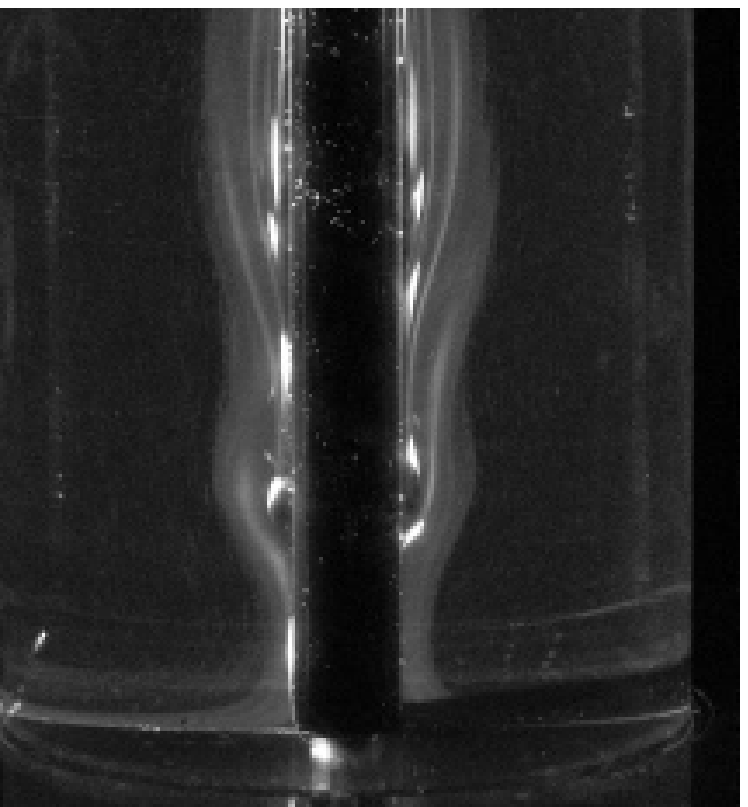}
\caption{Vortex breakdown for
$H/R=2.5$ and Re=2260. (a) without axial rod; (b) for $d = 5$ mm;
(c) for $d=10$ mm.}
\label{conbarra}
\end{figure}

We consider the situation in which an axial fixed rod of radius $d$ is
introduced. For $d = 1$ mm, the changes in the flow are very small in
comparison with the situation without rod.  However, for $d = 2.5$ mm
and $d = 5$ mm, we observe that the critical Reynolds number to
produce vortex breakdown is increased.  Experimental results are
summarized in Table~\ref{tab1}. In Figs.~\ref{diag} and \ref{rec} the
critical Reynolds number for the appearance of the first bubble as a
function of the aspect ratio and the radius of the rod respectively is
shown. We note that for values of ${\rm Re}$ larger that ${\rm Re}_c$,
the size of the bubble and its dynamics is still affected by the
presence of the inner cylinder.  For example, Figure \ref{conbarra},
shows vortex breakdown for $H/R=2.5$ and ${\rm Re}=2260$. We can
observe that without an axial rod, there are two oscillating bubbles
(Fig.  ~\ref{conbarra}a). For $d = 2.5$ mm, there are two VB, but, in
this case, they are steady (Fig.~\ref{conbarra}b) and for $d = 5$ mm,
only one VB appears. In addition, the size of the bubble corresponding
to the first VB was clearly decreased with the presence of the rods.

\section{Discussion}
\label{discussion}

From the tables and figures shown above it can be concluded that the
usage of the inner rods at rest, may be employed to control the vortex
onset. Comparing Figs.~\ref{Scritico} and \ref{rec}, we observe that
the experimental results about the effect of the cylinders on the
first bubble formation are in agreement with the prediction of the
simple model, assuming  the hypothesis that the swirl parameter
increases as ${\rm Re}$ increases.  This dependence of $S$ on ${\rm
Re}$ is suggested by the fact that both quantities are proportional to
the angular momentum of the flow. Recently, Husain \emph{et al}
proposed arguments that support this hypothesis \cite{Husain2003}.

The changes of the critical Reynolds number due to the rods presence
are clearly noticeable, as follows from table 1. We also
observe that in the presence of two bubbles, larger values of the
Reynolds number are needed for the development of the second bubble in
comparison with the situation without rod.  It is worth noting that
our results are in agreement with those of Hussain \emph{et al}
\cite{Husain2003}, in so far as that the very slender rod at rest does
not introduce significant changes in the flow. The radius ratio used
by these authors was $d/R=0.04$, which is similar to the small radius
ratio value we considered, that is $d/R=0.025$.  Significant results,
in our experiments, are obtained above $d/R=0.0625$.

It is also interesting to compare our results with those obtained
by Mullin \emph{et al} \cite{Mullin2000}. These authors reported 
that they did not obtain noticeable changes in the emergence of VB
with the addition of a inner cylinder with ratio $d/R=0.1$.
However, in our experiments we observed appreciable changes in the
flow, for values of the ratio $d/R=0.0625$ and $d/R=.125$.
Looking carefully at Figures 2 and 3 of  \cite{Mullin2000}, we
observe slight differences which, according to the authors, are
within the experimental errors. For this reason, we took specially
careful measurements in these cases.

\section{Summary and Conclusions}
\label{summary}

In this work, we present a method for controlling the onset of VB. It
consists basically of the addition of a small cylinder at rest in the
axis of the cylindrical container. The experiments we performed show
that this procedure increases the critical Reynolds number for the
emergence of VB, and consequently suppresses the onset of VB in a
certain range of values of Re.

The control technique proposed here is simpler than other previously
proposed in the literature, since it does not require additional
auxiliary devices. It is worth noting that our method, unlike the
approach proposed by Husain et al \cite{Husain2003}, does not imply
the addition of swirl near the axis. The simplicity of a method is in
general an interesting feature, becoming more feasible to be used in
engineering devices. Moreover, the required modification of the duct
is relatively small. The volume ratio (cylinder to row) is $V_1/V_2
\sim 4 \times 10 ^{-3}$, while the decrease of the critical Reynolds
number is about $10\%$ . So that the shift of the critical Reynolds
number is 20 times larger that the percent modification of the volume
of the cylinder, showing the effectiveness of the method.

The effect of the rods is consistent with the results of a simple
theoretical model of VB based on the failure of the quasi-cylindrical
approximation, which assumes that $S$ increases with increasing
Re. This agreement supports the idea of that to explain VB a local
description of the flow is adequate, and the details of the flows far
for the localization of VB do not play an important role. This point
is very appealing from a theoretical point of view, since it implies
that to study  VB in closed flows we must not consider all the
rather complex velocity fields, but that it is enough to consider only
the local field near the axis of the cylinder.

We acknowledge financial support from the Programa de Desarrollo
de Ciencias B\'asicas (PEDECIBA, Uruguay) and Grants FCE 9028 
and PDT54/037 (Conicyt, Uruguay).

\begin{table}
\caption{\label{tab1}Reynolds numbers corresponding to the appearance
of the first VB  for different aspect ratios and radius of the rod. }
\begin{center}
\begin{tabular}{|c||c|c|c|}  \hline
H/R   & d=1 mm &  d=2.5 mm  & d=5 mm     \\
\hline 1   & No VB & 892 & 1012
\\ \hline 1.5 & 940   & 1108  &  1132    \\ \hline 2   & 1300  &
1420  & 1636     \\ \hline 2.5 & 1756  & 1876  & 2260
\\ \hline
\end{tabular}
\end{center}
\end{table}
\bigskip


\newpage

\end{document}